\documentclass[twocolumn,showpacs]{revtex4}
\usepackage{amsmath}
\usepackage{amssymb}
\usepackage{bm}
\usepackage{epsfig}
\usepackage{graphicx}
\usepackage{color}

\renewcommand{\Re}{\mathop{\rm Re}}
\renewcommand{\Im}{\mathop{\rm Im}}

\begin{document}
\title{Second Harmonic Generation in Graphene}

\author{M.M. Glazov}\thanks{glazov@coherent.ioffe.ru}

\address{Ioffe Physical-Technical Institute of the Russian Academy of Sciences,
St. Petersburg, 194021 Russia}

\begin{abstract}
{We report on theoretical study of second harmonic generation in graphene. Phenomenological analysis based on symmetry arguments is carried out. It is demontrated, that in ideal graphene samples second harmonic generation is possible only if the radiation wave vector or its magnetic field is taken into account. Microscopic theory is developed for the classical regime of radiation interaction with electrons, where photon energy is much smaller than the charge carriers characteristic energy. It is demonstrated, that the emitted radiation can be strongly circularly polarized for the linearly polarized incident wave.}
\end{abstract}


\maketitle


\emph{Introduction.} Graphene, a monolayer of carbon atoms arranged in honeycomb lattice, became recently one of the most promising materials owing to its intriguing electronic and transport properties, for recent reviews see Refs.~\cite{Lozovik:2008eng,Morozov:2008eng,Falkovsky:2008eng,graphene_rmp} and references wherein. 

It is well known that \emph{non-linear} transport and optical phenomena are most sensitive to the fine details of energy spectrum and carrier scattering in condensed matter~\cite{ganichev_book,Fiebig:05}. One of the non-linear effects is the generation of \emph{dc} current under homogeneous excitation caused by the asymmetry of optical transitions and scattering in non-centrosymmetric media (photogalvanic effect) or by the transfer of photon momentum to the electron system (photon drag effect)~\cite{ivchenko05a}. The latter one does not require low symmetry of the structure, but it is masked by various photogalvanic effects in ordinary semiconductor quantum wells which lack an inversion center as a rule~\cite{ivchenko_ganichev}. By contrast, graphene is a centrosymmetric material which allowed unambiguous identification of the photon drag effect by its remarkable polarization dependence~\cite{karch2010}.

Here we consider theoretically light wave vector induced \emph{second harmonic generation} in graphene 
which consists in the formation of the \emph{ac} response: electric current or dielectric polarization oscillating at the double frequency as compared with the frequency of the incident radiation.  Experimentally second harmonic in graphene was observed in~\cite{dean:261910,PhysRevB.82.125411}, but microscopic theory is absent so far. Alike the photon drag effect, in centrosymmetric media like graphene second harmonic generation is allowed only provided that the photon momentum transfer to electron system is taken into account. This is in contrast to the third (and other odd) harmonic generation~\cite{2010arXiv1011.4841V,0953-8984-20-38-384204} as well as coherent injection of ballistic currents~\cite{PhysRevLett.105.097401,sun2010} which do not require photon momentum transfer, but appear in higher orders in incident electromagnetic field.

The second harmonic response is calculated here for low frequencies, e.g. terahertz radiation. In such case, the photon energy is much smaller than characteristic energy of electrons and a classical regime is realized for the radiation interaction with electrons in graphene, which brings about very strong polarization effects. In particular, second harmonic can be strongly circularly polarized for linearly polarized incident radiation.

\emph{Phenomenological theory.} We consider an ideal single layer graphene described by the point symmetry group $D_{6\rm h}$. This point group contains an inversion center, therefore, the second harmonic generation is possible only if one takes into consideration the light wave vector $\bm q$ transfer to electron ensemble or joint action of electric $\bm E$ and magnetic fields $\bm B$ of the incident radiation~\cite{Fiebig:05,ivchenko05a}. Since the magnetic field components $B_\alpha$ are proportional to certain combinations $q_\beta E_\gamma$, where greek subscripts enumerate Cartesian coordinates, in phenomenological considerations it is enough to allow for the contributions to the second order response being linear in the wave vector components and quadratic in the electric field amplitudes.

It is convenient to study the electric current, $\bm j(2\omega)$, induced at the double frequency $2\omega$ as a response to the incident radiation with the fundamental frequency~$\omega$. The electric current is related with the oscillating  polarization, $\bm P(2\omega)$, as $\bm j(2\omega)= -2\mathrm i \omega \bm P(2\omega)$. The general phenomenological expression describing the second harmonic generation reads
\begin{equation}
\label{phenom:gen}
j_\lambda(2\omega) = S_{\lambda\mu\nu\eta} q_\mu E_{\nu}E_{\eta}.
\end{equation}
Here $S_{\lambda\mu\nu\eta}$ is the material tensor and the electric field of the fundamental wave is taken in the complex form
\begin{equation}
\label{incident}
\bm E(\bm r, t) = \bm E e^{\mathrm i (\bm q\bm r - \omega t)} + {\rm c.c.}\ .
\end{equation}
The same complex form, Eq.~\eqref{incident}, is assumed for all other quantities in question: the magnetic field, $\bm B$, and the current density, $\bm j$.
It is worth mentioning that, in contrast to the photon drag effect where the \emph{dc} current is proportional to the bilinear combinations of electric field and its conjugate, $E_{\nu}E_{\eta}^*$, the second harmonic generation is described by the quadratic combinations $E_{\nu}E_{\eta}$. Hence, tensor $S_{\lambda\mu\nu\eta}$ is invariant with respect to permutations of the two last subscripts. The elements $S_{\lambda\mu\nu\eta}$ are, in general, complex, the presence of both real and imaginary parts describes the phase shift between the fundamental and second harmonics.

There are 7 independent parameters, $S_1\ldots S_7$, which describe second harmonic generation in the system of point group $D_{6\rm h}$:
\begin{subequations}
\label{phenom:7}
\begin{multline}
\label{phenom:7x}
j_x(2\omega) = S_1 q_x (E_x^2+ E_y^2) + S_2[q_x(E_x^2-E_y^2)+2q_yE_xE_y]\\
 + S_3 q_z E_xE_z +S_4 q_x E_z^2,
\end{multline}
\begin{multline}
\label{phenom:7y}
j_y(2\omega) = S_1 q_y (E_x^2+ E_y^2) + S_2[q_y(E_y^2-E_x^2)+2q_xE_xE_y]\\
 + S_3 q_z E_yE_z +S_4 q_y E_z^2,
\end{multline}
\begin{equation}
\label{phenom:7z}
j_z(2\omega) = S_5 q_z E_z^2 + S_6q_z(E_x^2+E_y^2) +S_7(q_xE_xE_z+q_yE_yE_z),
\end{equation}
\end{subequations}
where we assumed that the graphene layer occupies $(xy)$ plane.  Expressions~\eqref{phenom:7x} and \eqref{phenom:7y} are similar to phenomenological equations describing linear drag effect in graphene~\cite{2010arXiv1002.1047K}.

\begin{figure}[t]
\includegraphics[width=0.45\textwidth]{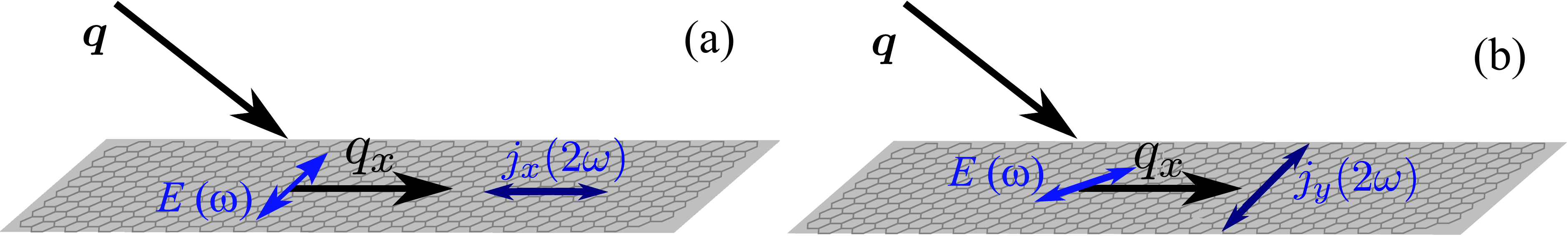}
\caption{Schematic illustration of the second harmonic generation geometry, Eqs.~\eqref{phenom:2}. It is seen that there are both longitudinal [panel (a)] and transverse [panel (b)] components of the second order response.}\label{fig:directions}
\end{figure}

The specifics of the graphene band structure imposes strong restrictions on values of the constants $S_1, \ldots, S_7$ in Eqs.~\eqref{phenom:7}. Indeed,  conduction and valence band of graphene are formed from $\pi$ orbitals of carbon atoms. As a result, the transitions in $z$ polarization are possible only with allowance for other, $\sigma$, orbitals which are separated from $\pi$ orbitals by several electronvolts. It implies that in the model where only $\pi$ orbitals are taken into account (e.g. in the classical frequency range of interest) constants $S_i=0$ for $i>2$ and second harmonic generation is described by the simplified expressions with only two independent parameters:
\begin{subequations}
\label{phenom:2}
\begin{equation}
\label{phenom:2x}
j_x(2\omega) = S_1 q_x (E_x^2+ E_y^2) + S_2[q_x(E_x^2-E_y^2)+2q_yE_xE_y],
\end{equation}
\begin{equation}
\label{phenom:2y}
j_y(2\omega) = S_1 q_y (E_x^2+ E_y^2) + S_2[q_y(E_y^2-E_x^2)+2q_xE_xE_y].
\end{equation}
\end{subequations}
Figure~\ref{fig:directions} schematically shows the geometry of the second harmonic generation and the response at a double frequency, $2\omega$. For instance, for $q_x\ne 0$, $q_y=0$, there is a component of the in-plane oscillating current $\bm j(2\omega)$ parallel to the light incidence plane described by $(S_1+S_2)q_x E_x^2+(S_1-S_2)q_xE_y^2$ [Fig.~\ref{fig:directions}(a)]. Additionally, there is a transverse contribution, $2S_2 q_x E_xE_y$, Fig.~\ref{fig:directions}(b).
Microscopic theory for these constants $S_1$ and $S_2$ is developed below.

Graphene samples are usually deposited on substrates, hence $z\to - z$ symmetry is broken. As a result, structure becomes non-centrosymmetric, described by the point group $C_{6\rm v}$, and second harmonic generation becomes possible for $\bm q=0$. It can be shown that additional contributions have the following form
\begin{subequations}
\label{phenom:substr}
\begin{equation}
\label{phenom:substrxy}
j_x(2\omega) =  S_3' E_xE_z, \quad 
j_y(2\omega) =  S_3' E_yE_z,
\end{equation}
\begin{equation}
\label{phenom:substrz}
j_z(2\omega) = S_5' E_z^2 + S_6' (E_x^2+E_y^2),
\end{equation}
\end{subequations}
where $S_3'$, $S_5'$ and $S_6'$ are some coefficients directly proportional to parameters $\partial_{15}^t$, $\partial_{31}^t$ and $\partial_{33}^t$ introduced in Eq.~(10) of Ref.~\cite{PhysRevB.82.125411}. As one can see from Eqs.~\eqref{phenom:substr} these contributions require transitions in $z$ polarization and are strongly suppressed in classical frequency range~\cite{2010arXiv1002.1047K} but they may take over for higher frequencies~\cite{PhysRevB.82.125411}.

\emph{Microscopic theory.} We consider here a doped graphene layer and assume that the characteristic electron energy (being the Fermi energy, $\varepsilon_{\rm F}$, for degenerate and the temperature, $k_{\rm B}T$, for non-degenerate electrons) is much larger than the photon energy $\hbar\omega$. This condition allows one to treat electron dynamics in the framework of Boltzmann equation. For typical carrier densities on the order of $10^{12}$~cm$^{-2}$ the electron Fermi energy $\varepsilon_{\rm F} \sim 100$~meV, therefore carriers are degenerate even at room temperature  and classical treatment is valid for microwave or terahertz radiation~\cite{karch2010}.

We introduce time, $t$, position,  $\bm r$, and momentum, $\bm p$, dependent   electron distribution function $f(\bm p, \bm r, t)$ which satisfies standard kinetic equation~\cite{karch2010,2010arXiv1002.1047K}
\begin{equation}
 \label{kinetic:gen}
\frac{\partial f}{\partial t} + \bm v_{\bm p} \frac{\partial f}{\partial \bm r} + {e}\left(\bm E + \frac{1}{c}[\bm v_{\bm p} \times \bm B]\right)\frac{\partial f}{\partial \bm p} = Q\{f\}.
\end{equation}
Here $\bm E$ and $\bm B$ are the electric and magnetic fields of the incident wave taken in the form Eq.~\eqref{incident}, $e$ is the electron charge, $e= -|e|$, $c$ is the speed of light, $\bm v_{\bm p}$ is the electron velocity in the given state $\bm p$,
\begin{equation}
 \label{vel:def}
\bm v_{\bm p} = \frac{\partial \varepsilon_{p}}{\partial \bm p} = v\frac{\bm p}{|\bm p|}.
\end{equation}
Here $\varepsilon_{p} = v |\bm p|$ is the electron dispersion, $v\approx c/300$ is effective electron speed in graphene, and $Q\{f\}$ in Eq.~\eqref{kinetic:gen} is the collision integral. In Eq.~\eqref{kinetic:gen} it is assumed that $f(\bm p, \bm r,t)$ describes the electron distribution in any of two equivalent degenerate valleys, $K$ and $K'$.
It is clearly seen from Eq.~\eqref{kinetic:gen} that electron motion is affected by the in-plane components of electric field $\bm E$ and by $z$ component of magnetic field $\bm B$.

The distribution function is sought in a form~\cite{2010arXiv1002.1047K}
\begin{multline}
\label{distrib}
f(\bm k, \bm r, t) = f_0(\varepsilon)+ [f_1^\omega(\bm k)e^{\mathrm i (\bm q\bm r - \omega t)}+ {\rm c.c.}] + \\
f_2^{0}(\bm k) + [f_2^{2\omega}(\bm k) e^{2\mathrm i (\bm q\bm r - \omega t)} + {\rm c.c.}]\ ,
\end{multline}
where $f_0(\varepsilon)$ is the equilibrium (Fermi-Dirac) distribution function, $f_1^\omega(\bm k)$ is the first order correction, and functions $f_2^{0}$, $f_2^{2\omega}$ are the second order corrections corresponding to the responses at zero frequency (photon drag effect) and at the frequency $2\omega$ (second harmonic), respectively. The terms containing higher frequencies (and, hence, higher powers of $\bm E$, $\bm B$) are disregarded here because they do not contribute to the second harmonic generation. The \emph{dc} response was calculated in Refs.~\cite{karch2010,2010arXiv1002.1047K}, below we find the function $f^{2\omega}_2(\bm k)$.

In order to determine $f^{2\omega}_2(\bm k)$ we solve Eq.~\eqref{kinetic:gen} by iterations. To that end one has to find first order correction, $f_1^\omega(\bm k)$, which can be expressed  up to the linear in $\bm q$ terms as~\cite{2010arXiv1002.1047K}
\begin{multline}
\label{f1k}
f_1(\bm k) = - {e\tau_{1,\omega}} f_0' \biggl\{(\bm E_0 \bm v_{\bm p})  -
\\
 \mathrm i \tau_{2,\omega}[ (\bm q \bm v_{\bm p})(\bm E_0 \bm v_{\bm p})-v^2(\bm q\bm E_0)/2] + 
\left.\frac{v^2(\bm q \bm E_0)}{2\omega}\right\}.
\end{multline}
Here we introduced $f_0' =  \mathrm d f_0/\mathrm d \varepsilon$,
\[
\tau_{n,\omega} = \frac{\tau_n}{1-\mathrm i \omega \tau_n}, \quad (n=1,2),
\]
where $\tau_1$ and $\tau_2$ are the relaxation times for the first and second angular harmonics of the distribution function which describe the momentum (first angular harmonic)  and momentum alignment (second angular harmonic) relaxation of electrons, respectively. These relaxation times can be determined in Born approximation as~\cite{2010arXiv1003.4731D}
\begin{eqnarray}
\label{taui}
& &\frac{1}{\tau_n} \equiv \frac{1}{\tau_n(\varepsilon_k)}= \\
 & &\frac{2\pi}{\hbar} n_{\rm imp} \sum_{\bm p} |V_{\bm k - \bm p}|^2 \frac{1+\cos{\vartheta}}{2} [1-\cos{(n\vartheta)}] \delta(\varepsilon_k - \varepsilon_p),\nonumber
\end{eqnarray}
where $n_{\rm imp}$ is the (two-dimensional) impurity density, $V_{\bm q}$ is the Fourier-transform of the impurity potential, $\vartheta$ is the scattering angle,  $(1+\cos{\vartheta})/2$ comes from the overlap of Bloch amplitudes and the normalization area is set to unity hereafter. Intervalley scattering processes are disregarded in Eq.~\eqref{taui}.

 The energy relaxation processes are neglected assuming that $\omega\tau_{\varepsilon} \gg 1$, $\tau_{\varepsilon}/\tau_{1,2} \gg 1$, where $\tau_{\varepsilon}$ is the energy relaxation time. 
 Equation~\eqref{f1k} is valid as long as $qv\tau_1,qv\tau_2,qv/\omega\ll 1$, that is if spatial variations of the distribution function occur on the scale exceeding by far the mean free path, this condition is readily fulfilled in the relevant frequency range. 

The second order correction satisfies the following equation
\begin{multline}
 \label{kinetic:2omega}
-2\mathrm i \omega f_{2}^{2\omega}(\bm p) + 2\mathrm i \bm q \bm v_{\bm p} f_{2}^{2\omega}(\bm p)+\\
 {e}\left(\bm E + \frac{1}{c}[\bm v_{\bm p} \times \bm B]\right)\frac{\partial f_1^\omega(\bm p)}{\partial \bm p} = Q\{f_{2}^{2\omega}\},
\end{multline}
and the electric current density is given by:
\begin{equation}
\label{j}
\bm j = 4e\sum_{\bm k} \bm v_{\bm p} f_{2}^{2\omega}(\bm p),
\end{equation}
where the factor $4$ takes into account spin and valley degeneracy. 
Solving Eq.~\eqref{kinetic:2omega}, retaining in $f_2^{2\omega}(\bm p)$ only linear in $\bm q$ or linear in $\bm B$ terms and expressing $\bm B = [(\bm q/q)\times \bm E]$ (for the sample in vacuum)
we arrive at the following expressions for the constants $S_1$ and $S_2$:
\begin{subequations}
\label{S}
\begin{eqnarray}
\label{S1}
& & S_1 = -\frac{e^3v^4}{2\omega}\sum_{\bm k} \tau_{1,\omega}f_0' \times \\
& & \left[ \frac{\tau_{1,2\omega}}{\varepsilon_k}(3+\mathrm i \omega\tau_{2,\omega}) + (1-\mathrm i \omega\tau_{2,\omega})\frac{\mathrm d\tau_{1,2\omega}} {\mathrm d\varepsilon_k} \right],\nonumber
\end{eqnarray}
\begin{eqnarray}
\label{S2}
& & S_2 = \frac{e^3v^4}{2\omega}\sum_{\bm k} \tau_{1,\omega}f_0' \times \\
& & \left\{ \frac{\tau_{1,2\omega}}{\varepsilon_k}(1+4\mathrm i \omega\tau_{2,2\omega}) - \frac{\mathrm d} {\mathrm d\varepsilon_k}[\tau_{1,2\omega}(1-2\mathrm i \omega\tau_{2,2\omega})] \right\}. \nonumber
\end{eqnarray}
\end{subequations}
It is noteworthy, that the physical mechanisms governing second harmonic generation are the same as in the photon drag effect~\cite{2010arXiv1002.1047K}. First one is related to the joint action of electric and magnetic fields of the incident radiation: electric field induces \emph{ac} current with the fundamental frequency $\omega$, while magnetic field oscillating at the same frequency induces both \emph{dc} current and the current component with the double frequency, $2\omega$. This is $EB$ mechanism of the second harmonic generation. Another $qE^2$ mechanism is connected with the spatial dependence of the electric field, its qualitative description can be carried out along the same lines as in Ref.~\cite{2010arXiv1002.1047K}. In quantum language the microscopic mechanisms governing these effects are due to the magneto-dipole and  quadrupole transitions, respectively.

\begin{figure}[t]
\includegraphics[width=0.45\linewidth]{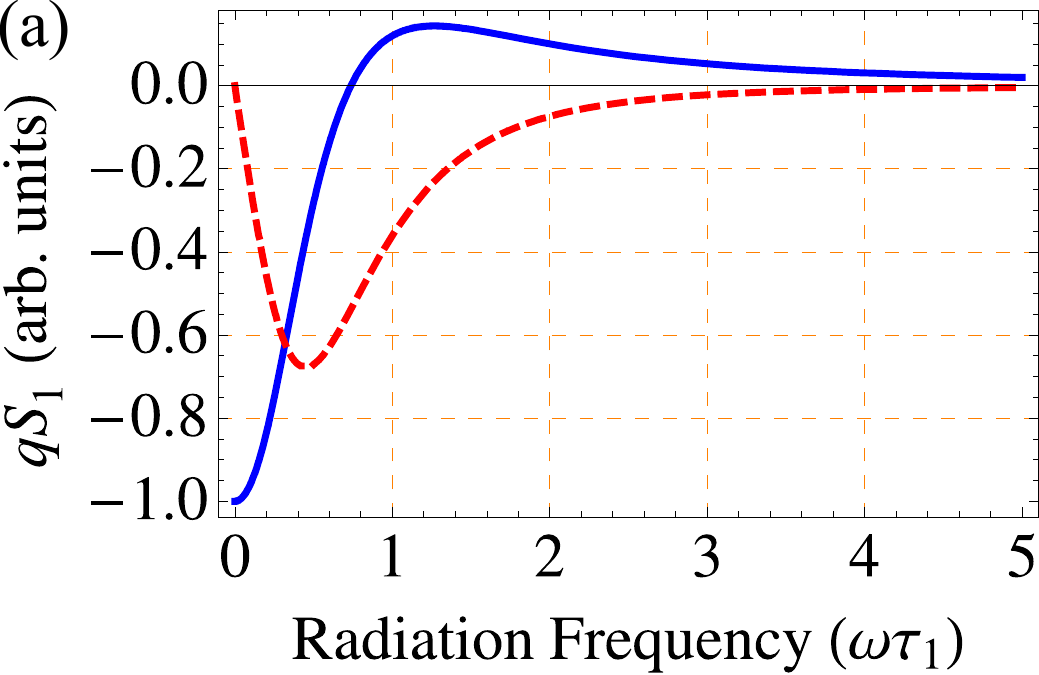}
\includegraphics[width=0.45\linewidth]{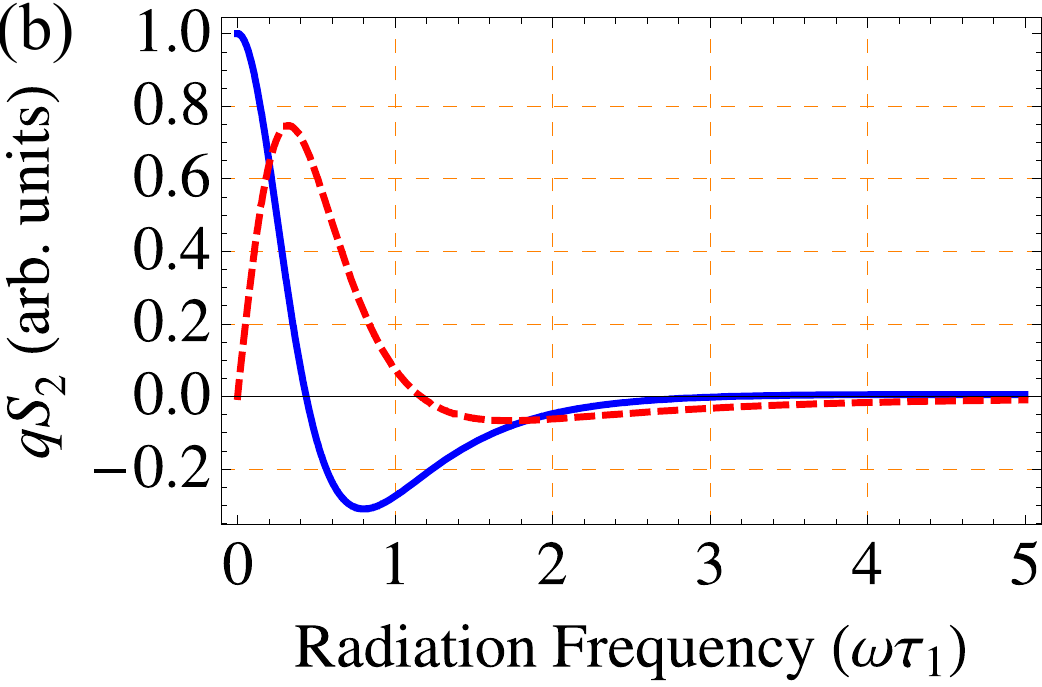}
\caption{Frequency dependence of the contributions to the second harmonic response: $qS_1$ is shown in panel (a), and $qS_2$ is shown in panel (b). Solid blue lines depict real part and dashed red lines depict imaginary part of the response. Values of current are given in arbitrary units, radiation frequency is presented in units of $\omega\tau_1$, where $\tau_1$ is momentum relaxation time. Calculation is carried out for the short-range scattering where $\tau_1=2\tau_2\propto \varepsilon_k^{-1}$ for degenerate electrons.} \label{fig:S1S2}
\end{figure}

Figure~\ref{fig:S1S2} shows frequency dependence of electric current components at double frequency $j(2\omega)$ proportional to coefficients $S_1$ [panel (a)] and $S_2$ [panel (b)], respectively. Calculation is carried out for the short-range scattering where $V_{\bm k - \bm p}$ in Eq.~\eqref{taui} is constant and relaxation times are interrelated as $\tau_1=2\tau_2\propto \varepsilon_k^{-1}$. The electron gas is assumed to be degenerate. It can be seen that the response at double frequency has complex non-monotonous behavior. In the static limit, $\omega\to 0$, the coefficients $S_1$ and $S_2$ are real and diverge as $1/\omega$, but the net current remains finite due to factors $\propto \bm q$ in Eqs.~\eqref{phenom:2}. Coefficients $S_1$ and $S_2$ become, up to common factor, equal to the constants $T_1$ and $T_2$ describing linear photon drag effect, see Eqs. (26) of Ref.~\cite{2010arXiv1002.1047K}, because at $\omega=0$ responses at zero and double frequencies are indistinguishable. At high frequencies, $\omega \tau_1 \gg 1$, $\omega \tau_2 \gg 1$, parameters $S_1$ and $S_2$ are proportional to $1/\omega^3$, hence, current density decays as $1/\omega^2$. The imaginary part of current is determined by the retardation between the driving electric field and electron velocity~\cite{karch2010,2010arXiv1002.1047K}. Therefore, it is zero in the limit of static fields, shows maximum at $\omega\tau_1\sim 1$ and decreases for $\omega\tau_1 \gg 1$. Typical values of $j(2\omega)$ for $\omega\tau_1 \sim 1$ are close to those of \emph{dc} current. For the experimental parameters of Ref.~\cite{karch2010}: $j(2\omega)/P\sim 0.5$~nAcm$/$W, where $P$ is the power density.


\emph{Polarization of the second harmonic}. The electromagnetic field emitted at the double frequency can be found by means of Maxwell equations. It is convenient to determine the vector potential $\bm A(\bm r, t)$ oscillating at $2\omega$, which satisfies the following equation~\cite{ll2_eng}
\begin{equation}
\label{Maxwell}
\Delta \bm A(\bm r, t) + \frac{4\omega^2}{c^2}\bm A(2\omega) = -\frac{4\pi}{c} \mathrm e^{2\mathrm i \bm q_\parallel \bm \rho - 2\mathrm i\omega t}\delta(z) \bm j(2\omega) + {\rm c.c.}
\end{equation}
where $\bm q_{\parallel}=(q_x,q_y)$ is the projection of the incident radiation wave vector onto the graphene layer and  we assumed that the sample is positioned in vacuum at $z=0$ plane.
Using one-dimensional Green's function~\cite{ivchenko05a}, the solution of Eq.~\eqref{Maxwell} can be written as
\begin{eqnarray}
\label{Maxwell:fin}
\bm A(\bm r, t) = \frac{\mathrm i \pi}{c q_z} \mathrm e^{2\mathrm i\bm q_\parallel \bm \rho - 2\mathrm i \omega t } \left[\bm j(2\omega)\mathrm e^{2\mathrm i q_z|z|} \right. \\
\left. -\bm e_z \frac{\bm j(2\omega)\bm q_\parallel}{q_z}(e^{2\mathrm i q_z|z|}-1) {\rm sign}{z} \right]+ {\rm c.c.} \nonumber
\end{eqnarray} 
Here $\bm e_z$ is the unit vector along $z$-axis. It follows from Eq.~\eqref{Maxwell:fin} that the emitted radiation propagates with the in-plane wave vector $2\bm q_\parallel$. The dependence of the vector potential on $z$ given by the factor $\exp{(2\mathrm i q_z|z|)}$ corresponds to the wave diverging from the quantum well.

The electric field of the emitted second harmonic is related with its vector potential by the standard expression $\bm E(2\omega) = -c^{-1}\partial \bm A(2\omega)/\partial t$. Making use of Eq.~\eqref{Maxwell:fin} one can see that the Stokes parameters of emitted radiation are given by
\begin{align}
\label{stokes}
P_l(2\omega) &= \frac{|j_x(2\omega)|^2-|j_y(2\omega)|^2}{|j_x(2\omega)|^2+|j_y(2\omega)|^2}, \nonumber
\\
P_l'(2\omega) &= \frac{2\Re\{j_x(2\omega)j_y^*(2\omega)\}}{|j_x(2\omega)|^2+|j_y(2\omega)|^2}, \nonumber \\
P_c(2\omega) &= -\frac{2\Im\{j_x(2\omega)j_y^*(2\omega)\}}{|j_x(2\omega)|^2+|j_y(2\omega)|^2}.
\end{align}
Here $P_l$ and $P_l'$ are the degrees of linear polarization in two coordinate frames: the $xy$ frame and in the $x'y'$ frame which is rotated by $45^{\rm o}$ with respect to $xy$ frame. Quantity $P_c$ is the circular polarization degree. In derivation of Eqs.~\eqref{stokes} we assumed small incidence angles, $q_\parallel \ll q_z$, and considered the emitted wave propagating along positive $z$ axis.	

\begin{figure}[t]
\includegraphics[width=0.45\linewidth]{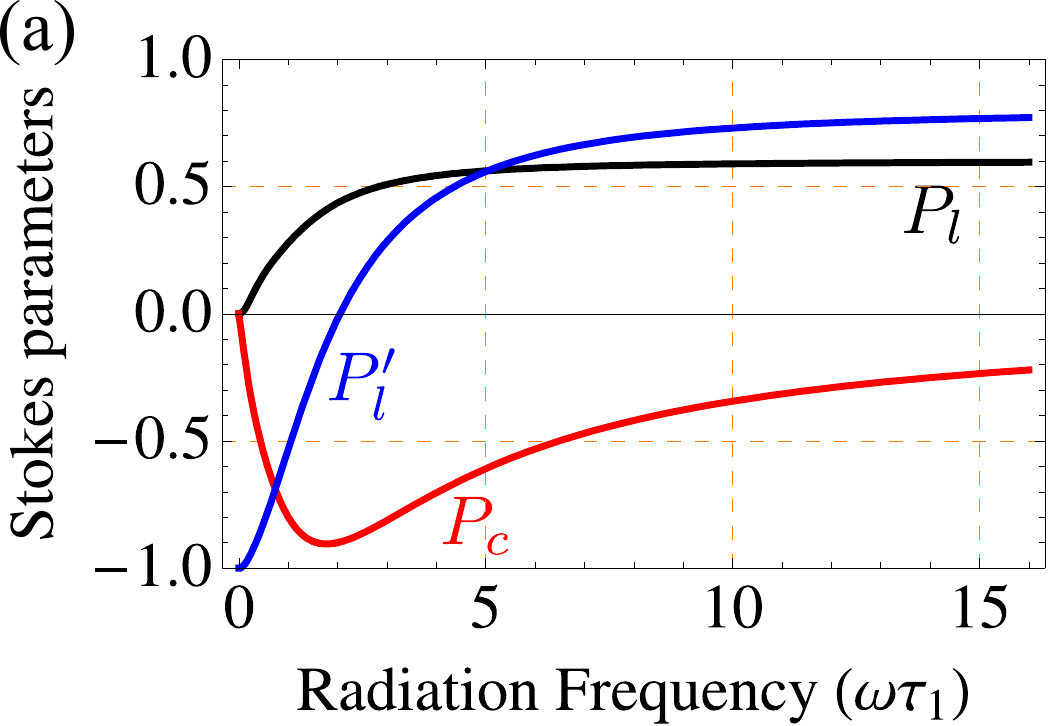}
\includegraphics[width=0.45\linewidth]{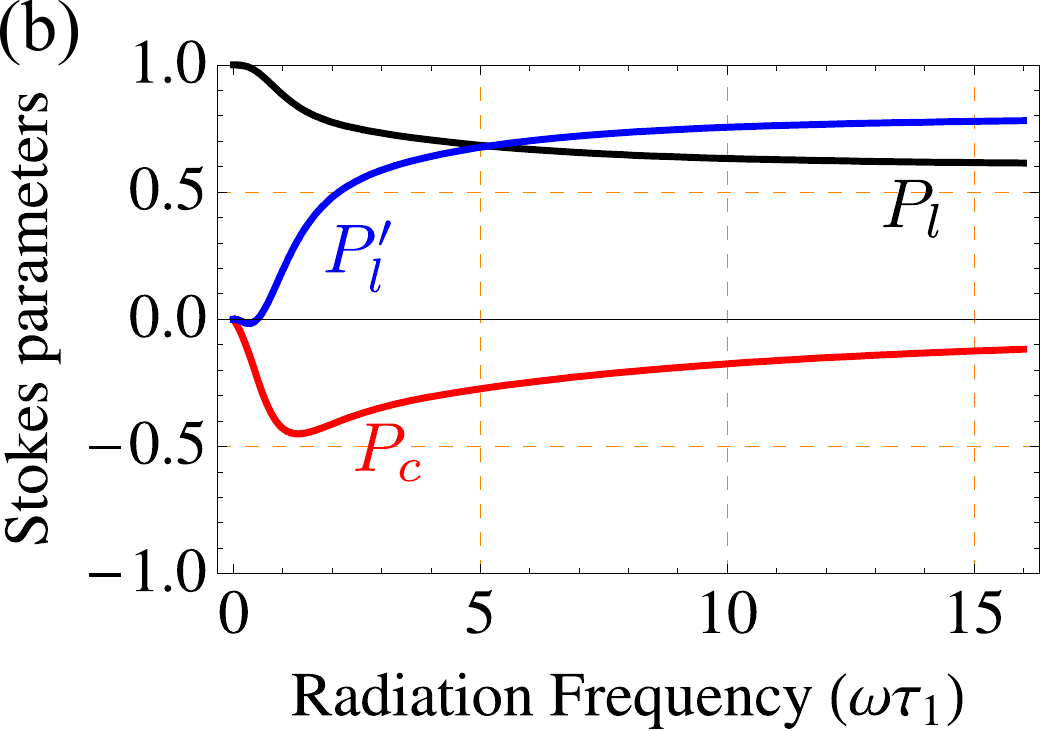}
\caption{Frequency dependence of Stokes parameters of second harmonic: $P_l$ is shown by black, $P_l'$ by blue and $P_c$ by red curves, respectively. It is assumed that the fundamental harmonic is fully linearly polarized in $x'y'$ axis frame: $|E_x|^2=|E_y|^2$, $2\Re\{E_xE_y^*\}=|E_x|^2+|E_y|^2$. Calculation is carried out for degenerate electrons in the model of the short-range scattering where $\tau_1=2\tau_2\propto \varepsilon_k^{-1}$ [panel (a)] and of the Coulomb scattering where $\tau_1=3\tau_2\propto \varepsilon_k$ [panel (b)].}\label{fig:stokes}
\end{figure}

We assume that the in-plane wave vector is directed along $x$ axis, $q_x\ne 0$, $q_y=0$. Non-trivial polarization effects take place if both $j_x(2\omega)$ and $j_y(2\omega)$ are non-zero. Such situation may arise for the linearly polarized incident radiation in $x'y'$ axis frame, where $|E_x|^2=|E_y|^2$, $2\Re\{E_xE_y^*\}=|E_x|^2+|E_y|^2$ [$P_l(\omega)=0$, $P_l'(\omega)=1$, $P_c(\omega)=0$] which is considered in the following. Stokes parameters for the second harmonic are plotted in Fig.~\ref{fig:stokes} as functions of the fundamental radiation frequency. Calculation presented in Fig.~\ref{fig:stokes}(a) is carried out for the short-range scattering where $\tau_1=2\tau_2\propto \varepsilon_k^{-1}$ [panel~(a)]. For comparison Fig.~\ref{fig:stokes}(b) shows calculation results for the unscreened Coulomb scattering potential, where $V_{\bm k - \bm p} \propto |\bm k - \bm p|^{-1}$ in Eq.~\eqref{taui} and one can show that  $\tau_1=3\tau_2\propto \varepsilon_k$. In all calculations it is assumed that the electron gas is degenerate.
 
Figure~\ref{fig:stokes} clearly demonstrates that all Stokes parameters are non-zero, in general. In the limit of static fields, $\omega\tau_1 = 0$, ellipticity is forbidden, $P_c(2\omega)=0$ in agreement with Fig.~\ref{fig:stokes}. In the limit of high frequencies $\omega\tau_1, \omega\tau_2 \gg 1$ (but $\hbar\omega \ll \varepsilon_{\rm F}$) constants $S_1$ and $S_2$ are related as $S_2=S_1/2$ which yields
\[
P_l = \frac{3}{5}, \quad P_l'=\frac{4}{5}, \quad P_c =0, \quad \mbox{ at} \quad \omega\tau_1,\omega\tau_2 \gg 1.
\]

It follows from Fig.~\ref{fig:stokes} that second harmonic radiation is, in general, elliptically polarized. The remarkable degree of circular polarization appears at $\omega\tau_1 \sim 1$. For the short-range scattering considered in panel (a) of Fig.~\ref{fig:stokes} the maximum absolute value of the circular polarization $|P_c|$ is about 90\% at $\omega\tau_1 \approx 1.78$. The maximum value of the circular polarization degree for the Coulomb scattering is somewhat smaller, $|P_c| \approx 45$\% at $\omega\tau_1 \approx 1.31$.  Physically, the circularly polarized second harmonic emission is a result of retardation: $j_x(2\omega)$ and $j_y(2\omega)$ are phase shifted with respect to each other. This phase difference can be significant resulting in almost fully circularly polarized second harmonic emission. In agreement with qualitative considerations and Fig.~\ref{fig:S1S2}, the phase shift vanishes both for static and high frequency fields, where $S_1$ and $S_2$ are real. Circular polarization approaches zero in these two limits.

The appearance of the linear polarization in $xy$ axis frame, $P_l(2\omega) \ne 0$, and of the circular polarization, $P_c(2\omega) \ne 0$ can be considered as a non-linear polarization conversion process. Inversion of initial linear polarization $P_l'(\omega) \to -1$ results in the reversal of the circular polarization, $P_c$, and of the linear polarization in $x'y'$ axis frame, $P_l'(2\omega)$. Linear polarization in $xy$ axis frame does not change its sign.


\emph{Concluding remarks}. We have developed theory of second harmonic generation in graphene. The analysis shows that in the simplest model of  conduction and valence bands formed by $\pi$ orbitals of carbon atoms, the response at the double frequency appears with allowance for the photon wave vector transfer to electron system and it is described by two independent parameters. The microscopic theory is presented here for the classical frequency range, where the photon energy is much smaller than the characteristic energy of charge carriers in graphene. In this regime effects of other orbitals are negligible and the second harmonic generation is fully determined by the specifics of the radiation momentum transfer to ``Dirac'' electrons in graphene. 

We have analyzed the polarization of the emitted radiation. It is demonstrated, that even for linearly polarized fundamental harmonic, the second harmonic can be strongly circularly polarized. The degree of circular polarization can be controlled by the radiation frequency, it can be reversed by changing the linear polarization of incident wave.


Author is grateful to E.L. Ivchenko and S.A.~Tarasenko for valuable discussions. The support by RFBR and ``Dynasty'' Foundation---ICFPM is gratefully acknowledged.


\begin{thebibliography}{99}

\bibitem{Lozovik:2008eng} Y.~E. Lozovik, S.~P. Merkulova, A.~A. Sokolik, Physics-Uspekhi \textbf{51}, 727 (2008).
%
\bibitem{Morozov:2008eng} S.~V. Morozov, K.~S. Novoselov, A.~K. Geim, {Physics-Uspekhi} \textbf{51}, 744, (2008).
%
\bibitem{Falkovsky:2008eng} L.~A. Falkovsky, {Physics-Uspekhi} \textbf{51}, 887 (2008).
%
\bibitem{graphene_rmp} A. H. Castro Neto, F. Guinea, N. M. R. Peres, 
et al.,
{Rev. Mod. Phys.} \textbf{81}, 109 (2009).
%
\bibitem{ganichev_book} S.~D.~Ganichev, W. Prettl, \textit{Intense Terahertz Excitation of Semiconductors}, {Oxford, UK} (2006).

\bibitem{Fiebig:05}%
  
  {%
   {{M.}~{Fiebig}}, 
  {{V.~V.} {Pavlov}},
  {{R.~V.} {Pisarev}}, 
{J. Opt. Soc. Am. B} 
 \textbf{{22}}, {96} (2005).
 

%
\bibitem{ivchenko05a} E.~L. Ivchenko, \textit{Optical Spectroscopy of Semiconductor Nanostructures}, {Alpha Science, Harrow UK} (2005).
%
\bibitem{ivchenko_ganichev} E.~L.~Ivchenko, S.~Ganichev, in \textit{Spin physics in semiconductors}, ed. M.I. Dyakonov,  {Springer-Verlag} (2008).
%
\bibitem{karch2010}%
   {{J.}~{Karch}}, 
  {{P.}~{Olbrich}}, 
  {{M.}~{Schmalzbauer}}, 
et al.,
{Phys. Rev. Lett.} \textbf{105}, {227402} (2010).

\bibitem{dean:261910}%
  
  {%
   {{J.~J.}\ {Dean}}\ and\ {{H.~M.}\ {van Driel}},\ }%
{%
{Appl. Phys. Lett.}\ }%
  \textbf{{95}}, {261910} ({2009}).
  
  
\bibitem{PhysRevB.82.125411}%
  
  {%
   {{J.~J.}\ {Dean}}\ and\ {{H.~M.}\ {van Driel}},\ }%
{%
  {Phys. Rev. B}
  \textbf{{82}}, {125411} ({2010}).
  
  

 
\bibitem{2010arXiv1011.4841V}%
   {%
   {{F.~T.}\ {{Vasko}}},\ }%
preprint arXiv:1011.4841
 (2010).
 
 
\bibitem{0953-8984-20-38-384204}%
  
  {%
   {{S.~A.}\ {Mikhailov}}\ and\
   {{K.}~{Ziegler}},\ }%
{%
{Journal of Phys.: Cond. Mat.}\ }%
  \textbf{{20}}, {384204} ({2008}).
  
  
\bibitem{PhysRevLett.105.097401}%
  
   {{E.}~{Hendry}}, 
  {{P.~J.}\ {Hale}}, 
  {{J.}~{Moger}}, 
  et al.,
{{Phys. Rev. Lett.}}\
  }%
  \textbf{{105}}, {097401} ({2010}).
  
  
\bibitem{sun2010}%
  
   {{D.}~{Sun}}, 
  {{C.}~{Divin}}, 
  {{J.}~{Rioux}}, 
  et al.,
{%
{Nano Letters}} \textbf{{10}}, {1293} (2010). 
  
\bibitem{2010arXiv1002.1047K}%
  
   {{J.}~{{Karch}}}, {{P.}~{{Olbrich}}}, 
  {{M.}~{{Schmalzbauer}}}, 
et al.,
preprint arXiv:1002.1047  (2010).


\bibitem{2010arXiv1003.4731D}%
  
  {%
   {{S.}~{{Das Sarma}}}, {{S.}~{{Adam}}}, 
  {{E.~H.}\ {{Hwang}}},\ and\ 
  {{E.}~{{Rossi}}},\ }%
preprint arXiv:1003.4731 (2010).

\bibitem{ll2_eng}%
  
  {%
   {{L.D.}~{Landau}}\ and\  {{E.M.}~{Lifshitz}},\ }%
  \emph{{The Classical Theory of Fields}}, {Butterworth-Heinemann, Oxford UK}, (1975).}
\end{thebibliography}


%
\end{document}